# Local phase delay effect on the asymmetric spectroscopy of plasmon-exciton coupling systems

Aiqin Hu,[1] Weidong Zhang,[1] Lulu Ye,[1] Ying Gu,[1,2,3] Zhaohang Xue,[1] Hai Lin,[1] Jinglin Tang,[1] Qihuang Gong,[1,2,3] and Guowei Lu[1,2,3,*]

[1] State Key Laboratory for Mesoscopic Physics, Frontiers Science Center for Nano-optoelectronics & Collaborative Innovation Center of Quantum Matter, School of Physics, Peking University, Beijing 100871, China.
[2] Collaborative Innovation Center of Extreme Optics, Shanxi University, Taiyuan, Shanxi 030006, China.
[3] Peking University Yangtze Delta Institute of Optoelectronics, Nantong 226010, Jiangsu, China.
Email: guowei.lu@pku.edu.cn

**Abstract**

The phase delay of a local electric field, being well-known in plasmonic nanostructures, has seldom been investigated to modulate the plasmon-exciton interaction. Here, with the single-particle spectroscopy method, we experimentally investigate the phase effect in plasmon-exciton coupling systems consisting of monolayer $WSe_2$ and an individual gold nanorod. The local plasmon phase delay is tuned by adopting various nanorods with different resonant energies respective to the exciton. We find that the local plasmon phase delay between the excitons and the plasmonic modes is as equally essential as the amplitude. The phase delay modulates the plasmon-exciton coupling considerably, resulting in an asymmetric spectral line-shape due to the interference behavior. There is an excellent agreement for the phase delay between the numerically calculated near-field phase distribution and the experimental results. The local phase delay can act as an effective way to modulate the properties of plexcitonic coupling at the nanoscale, which may have potential applications in nanoscale sensing, solar energy devices, and enhancing nonlinear processes.

Understanding the delicate nature of light-matter interactions at a nanometer scale is the central issue in nanophotonics. Plasmon-exciton coupling remains an active subject of fundamental and applied research in the field of nano-plasmonics, and it has garnered much interest [1-12]. Many observations of intriguing phenomena in the weak, medium, and strong coupling regimes have been reported, such as plasmon-enhanced fluorescence [1], plasmon-enhanced absorption [13], plasmonic quenching, Fano resonances [14], and Rabi splitting [1, 8, 15, 16]. Much attention has been paid to the coupling strength, composition, and dispersion of a plasmon-exciton hybrid [17-19]. The coupling coefficient can be controlled by tuning the detuning, such as adjusting the plasmon resonance frequencies via changing the geometry [4] of the nanoparticles or the ambient index [1, 2], tuning the exciton resonance energy through altering the layers of transition metal dichalcogenides (TMDs) [5] or the ambient temperature [3]. These efforts to understand the physical micro-process are essential for developing novel sensing, solar energy devices, and enhancing nonlinear processes.

It is well-known that both the amplitude and the phase can strongly modulate the mode coupling process. For instance, the studies about the Rabi splitting and the Fano resonances of plasmon-exciton systems have often been investigated regardless of the absorbance of the excitons. By contrast, Ding et al. reported a novel finding for an asymmetric spectroscopic regime for both the Rabi splitting and the transparency dip. The asymmetric nature was inherently tied to the non-negligible absorbance of the excitons and the substantial interference-induced energy repartitioning of the resonance peaks [20]. The phase had to be equally as essential as the amplitude to modulate the coupling spectral line-shape [21]. For plasmonic nanostructures, the local phase depends strongly on space and resonance conditions, which have been successfully used to engineer the local field of the optical components in the metasurface field. However, the effect of the local phase delay has not been studied in plasmon-exciton coupling systems.

This study investigates the local plasmon phase effect resulting in an interference-induced asymmetrical spectra line-shape. We demonstrate theoretically the plasmon phase delay effect on the spectral line-shape using a semi-classical coupling model [22, 23]. In the experiment, we investigate the plasmon-exciton hybrid system consisting of a gold nanorod (GNR) and monolayer $WSe_2$ with the single-particle spectroscopy method. The hybrid is assembled using the atomic force microscope (AFM) nanomanipulation technique. We observe the asymmetrical scattering spectra of the hybrid due to the local phase delay between the plasmon and exciton. The blue detuning between the localized surface plasmon resonance (LSPR) modes and the exciton can induce an apparent out-of-phase interference and modulate the spectral line-shape significantly. We calculate the spatial distribution intensity and the phase of the near field with the finite-difference time-domain (FDTD) method. There is an excellent agreement between the theoretical and experimental results for the coupling spectra with different detuning. Our work brings a deep microscopic understanding of the phase effect in a plasmon-exciton system, which has potential applications in nanoscale quantum sensing, solar energy devices, and enhancing nonlinear processes.

The plasmon-exciton coupling system is shown schematically in Fig. 1(a). The LSPR mode of the GNR and the exciton of the monolayer $WSe_2$ are described as two modes with a given resonance frequency ($\omega_p/\omega_e$) and loss rate ($\gamma_p/\gamma_e$). For the $WSe_2$, $\omega_e$ and $\gamma_e$ are nearly fixed, and they can be determined experimentally using photoluminescence (PL). Additionally, $\gamma_p$ is about 3–5 times $\gamma_e$ according to the experimental data [24]. The Hamiltonian of the plasmon mode and the exciton is $H_c = \omega_p a^+ a + \omega_e b^+ b$. The interaction between the two modes is $H_{int} = g(a^+ + a)(b^+ + b)$. Using the rotating-wave approximation, $H_{int} = g(a^+ b + ab^+)$. The interaction between the incident light and the modes can be written as $H_{light} = \sqrt{\eta_p} E_0 (a^+ e^{-i\omega t} + a e^{i\omega t}) + \sqrt{\eta_e} E_e (b^+ e^{-i\omega t} + b e^{i\omega t})$. Here, $\eta$ is the coupling rate between the electric field and the modes, and it is different for the two modes. $E_0$ and $E_e$ are the electric fields felt by the plasmon and the excitons, respectively. Then the dynamics of the modes can be written as $\dot{a} = i[H, a] - \gamma_p a$ and $\dot{b} = i[H, b] - \gamma_e b$. Using the input-output relation, we can obtain the scattering spectrum using the solution of the dynamics equations with $I_{sc} \propto C_{nf} <a^+ a>$. $C_{nf}$ is a transfer coefficient from the near-field to the far-field [25]. The calculated scattering spectrum only contains the contribution of the plasmon mode [22].

The local phase shift between the scattering near-field of the plasmonic resonator and the incident radiation changes across the resonance [21]. We calculate the phase and the strength of the scattering light of a GNR, as shown in Fig. 1(b and c). The phase and the strength of the plasmonic near-fields of the antennas are both position-dependent and wavelength-dependent. In the present study, the frequency of the exciton is fixed. This provides a unique mechanism for interference control on a subwavelength scale through tuning the antenna resonant wavelength crossing the exciton frequency. In the GNR-WSe$_2$ coupling system, we consider that the excitons are excited mainly by the plasmonic near-field and not by the external field since the plasmonic enhancement effect dominates the near-field. Furthermore, as shown in Fig. 1(a), the plasmonic mode is excited mainly by the external field based on the assumption that the excitons' scattering field is negligible. Therefore, there is a phase difference when the excitons couple to the plasmon modes.

Here, we investigate the spectral line-shape in three cases: $E_e = 0$, $E_e = f_2$, and $E_e = f_2 e^{i\theta}$, where $f_2$ is the amplitude enhancement ratio of $E_e$ to $E_0$ and $\theta$ is the phase difference between $E_e$ and $E_0$. The calculated results are shown in Fig. S1 and Fig. 1(d). As shown in Fig. 1(d), the spectral characteristics change significantly with various $E_e$. When $E_e = 0$, corresponding to the Fano resonance, the calculated spectrum shows symmetric features. When $E_e \neq 0$, the spectra convert to the asymmetric features. Similar results have been reported, representing coupling systems modified by the excitonic transition [18, 20]. Notably, phase $\theta$ can definitely modulate the spectral features like the Born-Kuhn configurations. To show the influence of the out-of-phase interference behavior between the plasmons and excitons, we calculate the hybrid's spectra with different phases, as shown in Fig. 1(e). The results indicate that the line-shape of the two-hybrid resonance modes is modulated strongly by the phase. The interference process induces the spectral shape reforming by repartitioning the weight of the two split peaks.

In the following, we perform control experiments to demonstrate the phase effect. Since the monolayer TMDs possess a considerable transition dipole moment $\mu_e$ and the plasmonic nanoparticles provide a strong mode confinement $V$ [26], the hybrid facilitates strong nanoscale light-matter interaction according to $g \propto \mu_e/\sqrt{V}$ [27, 28]. Hence, the combination of the monolayer TMDs and the metal nanoparticles represents an ideal platform for investigating plasmon-exciton interaction at room temperature. In the experiment, the plasmon-exciton hybrid consists of monolayer WSe$_2$ and an individual GNR (details shown in Fig. S2). The sample is investigated with the single-particle spectroscopy method (details are shown in

Fig. S3) [24, 29, 30]. The monolayer WSe$_2$ can be confirmed using the AFM topography and the PL spectrum (Fig. S3(b and c)). Then we manipulate a single GNR elaborately near the WSe$_2$ sheet boundary using the nanomanipulation method. The step-by-step representative manipulation process is shown in Fig. 2(a) (e.g., for the GNR with a resonance frequency at the wavelength of ~ 723 nm and the linewidth ~149 meV). Fig. 2(b) shows the distance-dependent mapping of the plasmon-exciton coupling spectra. The spectral evolution is partly due to the variation of the coupling strength. As shown in Fig. 2(c), typical spectra measured at three particular positions (i.e., position 1 (or 3) indicates the GNR on the glass (or WSe$_2$), and position 2 indicates the GNR on the boundary of WSe$_2$). These results imply that the observable coupling occurs once the GNR is onto or near the monolayer WSe$_2$. We fit the experimental results using the semi-classical model with the amplitude ratio $f_2$ and phase $\theta$ of the parameters. As shown in Fig. S4, $f_2$ and phase $\theta$ are dependent on the separation between the GNR and the WSe$_2$. This result is consistent with the fact that both the phase and the strength of the GNR are position-dependent. However, the distance-dependent experiment mixes several uncontrollable factors. For example, the orientation of the GNR relative to the boundary of the WSe$_2$ also influences the coupling process. Therefore, we adjust the local plasmon phases by manipulating various GNRs with different resonant frequencies onto the WSe$_2$ (Fig. 2(d and e), Fig. S5). It should be noted that the resonant frequency of the GNR is redshifted when it is moved onto the WSe$_2$ from the glass [24].

We compare and analyze the scattering spectra of the hybrids after the GNR is wholly moved onto the WSe$_2$ shown in Fig. 2(f) to reduce the influence of the distance and orientation. Here, we focus on three coupling spectra types when the detuning $\Delta<0$, $\Delta\approx0$, and $\Delta>0$. As shown in Fig. 3(a), for detuning $\Delta<0$, when $f_2=0$ & $\theta=0$ (yellow dashed line) or $f_2\neq0$ & $\theta=0$ (green dashed line), the spectral dip and the low-energy branch cannot be fitted well. Only when $f_2\neq0$ & $\theta\neq0$ (black line), the fitted results and the experimental spectra show a good agreement. When the detuning is close to zero ($\Delta\approx0$), the spectral line-shapes can both be fitted by the models where $f_2\neq0$ & $\theta=0$ and $f_2\neq0$ & $\theta\neq0$, as shown in Fig. 3(b). However, when the detuning is large, the spectral line-shape can only be fitted well by the model where $f_2\neq0$ & $\theta\neq0$ (Fig. 3(a) and 3(c)). We fit a sufficient amount of the scattering spectra of the hybrids for the GNRs with different LSPRs using the model with $f_2\neq0$ & $\theta\neq0$ (the detailed fitting parameters, including the phase, resonance frequencies, and loss rates, are shown in Tables S1–S3 and Fig. S6). The fitting phase increases (from ~ 0.5 π to 1.1 π) as the LSPR wavelengths vary from blue to red detuning. We note that the coupling strength satisfies the strong coupling criterion according to the scattering spectra mentioned above. Furthermore, we perform the same measurements using the GNRs with a larger linewidth (180 ± 20 meV), which leads to a weaker coupling strength. The results also demonstrate the phase delay effect in the weak coupling regime, as shown in Fig. S7. Therefore, the coupling configurations where $f_2{\neq}0$ & $\theta{\neq}0$ can be applied well to all situations. This suggests that the phase effect modulates the plasmon-exciton coupling considerably, resulting in the asymmetric spectral line-shape due to the interference behavior. This phase control can provide a way to tune the energy repartitioning and the correlated electronic state occupations.

The localized plasmon is a well-known platform for controlling the electric field at the nanoscale [31]. The strength of the external light field felt by the WSe$_2$ is far less than the $E$-near field of the GNR. The phase difference is roughly equal to the local plasmon delay near the GNR. To verify the assumption, we investigate the $E$-near field felt by the WSe$_2$ using the FDTD simulation numerically. The $E$-near field is shown in Fig. 4(a) and 4(b). The $E$-near field at the interface is extracted, and there is an apparent spatial distribution for the strength and the phase [32-34]. The $E$-near field is strong at the hot-spot, and the phase is significantly variable, as shown in Fig. 4(c). We integrate the $E$-near field around the GNR with an area of about 240 nm × 460 nm to represent the averaged driving field for the WSe$_2$. The near-field's integral value has a noticeable phase difference respective to that of the external illumination light. We calculate the GNRs with different resonant frequencies and extract the $E$-near field at the resonant energy of exciton. We obtain the strength and phase of the integral value as a function of the resonance energy, as shown by the solid lines in Fig. 4(d) and 4(e). We find that the FDTD simulation results show a good agreement with the experimental results by fitting with the semi-classical model. When the detuning $\Delta<0$, the plasmon phase delay is about 0.5π, resulting in more considerable deconstructive interference. This can explain why the phase effect shows a pronounced influence when the detuning $\Delta<0$ (Fig. 3(a)). Furthermore, the local phase of the hot spot can be modulated efficiently through the plasmonic multi-mode coupling (the details are shown in Fig. S8, Supplementary Materials). This provides a way to adjust the energy distributions in the plasmon-exciton coupling system by controlling the local phase.

**Conclusions**

In conclusion, we demonstrate the plasmon phase effect in the plasmon-exciton coupling and interference-induced asymmetrical line-shape with different parameter configurations. Based on the semi-classical coupling model, we find that both the excitation of the excitons and the phase delay between the excitons and the plasmonic modes should be considered in the plasmon-exciton coupling. We investigate the plasmon-exciton hybrids consisting of monolayer WSe2 and an individual GNR to verify the phase effect in the experiment. We obtain the scattering spectra of the hybrid with the single-particle spectroscopy method. The coupling configurations with the phase delay can fit all experimental coupling spectra. There is an excellent agreement between the numerical simulations and experimental results for the phase delay. These results demonstrate that the phase control can provide a way to tune the plasmon-exciton coupling process. Our study contributes to a deep understanding, and it enriches microscopic scale understanding of plasmon-exciton hybrid systems substantially.

**Conflicts of interest**

The authors declare no competing interests.

**Acknowledgments**

This work was supported by the National Key Research and Development Program of China (Grant No. 2018YFB2200401), the Guangdong Major Project of Basic and Applied Basic Research (Grant No. 2020B0301030009), and the National Natural Science Foundation of China (Grant Nos. 91950111, 61521004, and 11527901).

**Figures**

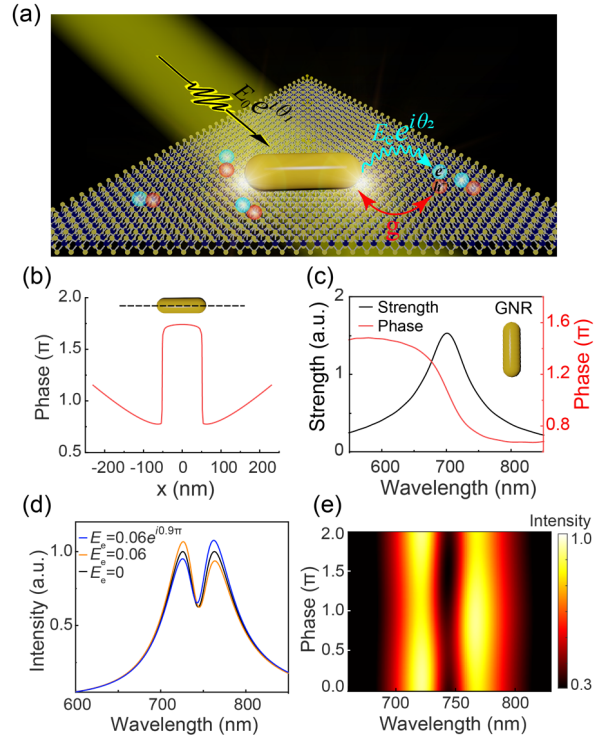

**Figure 1**. (a) A schematic to describe the physical processes. (b) Calculated relative phase at different positions across a GNR antenna longitudinal axis. (c) Calculated relative strength (black) and local phase (red) of scattered light near a GNR antenna. (d) Theoretically calculated scattering spectra based on the model when $E_e = f_2 e^{i\theta} = 0$ (black line), $E_e = 0.06$ (orange line), and $E_e = 0.06 e^{i0.9\pi}$ (magenta line). The resonance wavelength/loss rate of the LSPR mode (exciton mode) is at 744 nm (744 nm)/160 meV (48 meV). The coupling coefficient is 0.03 eV. (e) Normalized scattering spectra for the model with different phases.

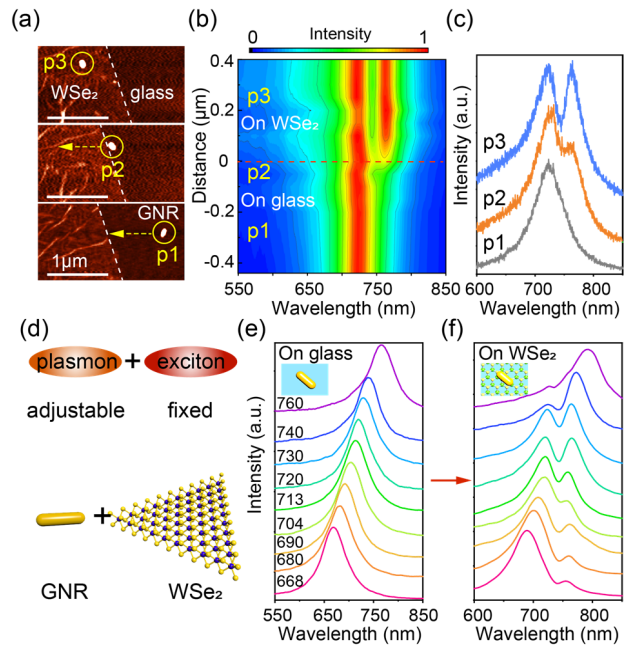

**Figure 2.** (a) AFM images demonstrating the nanomanipulation process. The white dashed line indicates the WSe$_2$ sheet boundary (scale bar of 1 μm). The yellow circles highlight the positions of the GNR. (b) The mapping indicates the distance-dependence scattering spectra of the hybrid. The positions labeled as "p1", "p2", and "p3" indicate the GNR on the glass substrate, near the boundary, and on the monolayer WSe$_2$. (c) Typical scattering spectra when the GNR is at the positions "p1" (gray), "p2" (orange), and "p3" (blue). (d) Schematic for tuning the resonant plasmon frequencies to couple with the exciton. (e) Experimental scattering spectra of various individual GNR on the glass substrate. The linewidth of the measured scattering spectra is 130±20 meV. (f) Experimental scattering spectra of the hybrid corresponding to the GNRs in (e) before coupling to the monolayer WSe$_2$.

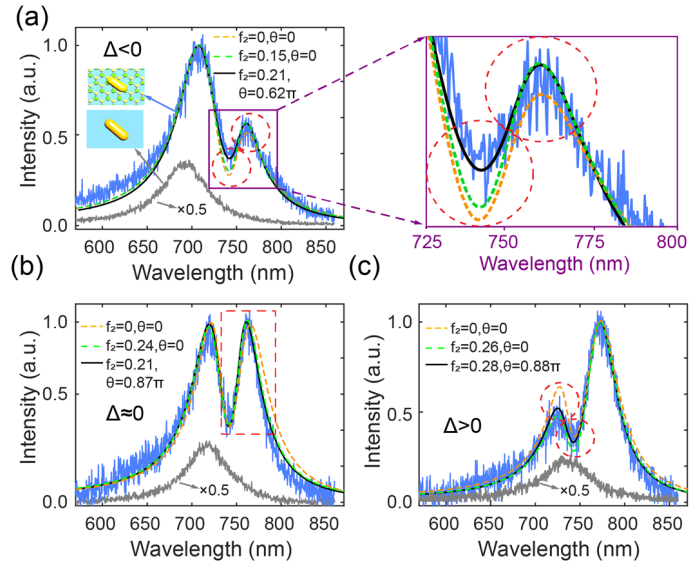

**Figure 3.** (a) Left column: Experimental scattering spectra of GNR-WSe$_2$ hybrids with detuning $\Delta<0$ (blue line), and individual GNR on the glass substrate (gray line). Different fitting results using the semi-classical coupling model when $f_2=0$ & $\theta=0$ (yellow dashed line), $f_2\neq0$ & $\theta=0$ (green dashed line), and $f_2\neq0$ & $\theta\neq0$ (black line). Right column: Enlarged view of (a) to show the fitting details highlighted by the red dashed circles. (b), (c) Experimental scattering spectra and fitted results of hybrids with detuning $\Delta\approx0$ (b), and $\Delta>0$ (c).

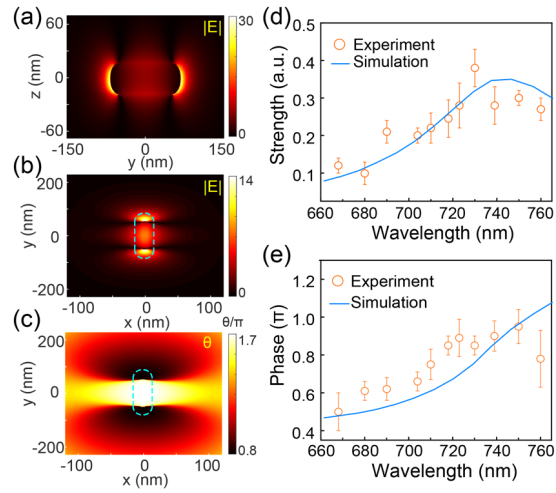

**Figure 4.** (a) The electric field of a GNR at the *y-z* component plane (the diameter is 30 nm, and the length is 140 nm) with the resonance wavelength at ~710 nm. (b), (c) Relative strength and phase distribution near the GNR bottom surface. The cyan dashed line highlights the geometry of the GNR. The relative strength (d) and phase (e) of $E_e$ at the resonant energy of excitons for the GNRs with different resonant wavelengths in the simulations (blue line) and the experiments (orange circles).

# Supplemental Materials

# Revealing Phase Effect in Plasmon-Exciton Coupling

Aiqin Hu,[1] Weidong Zhang,[1] Lulu Ye,[1] Zhaohang Xue,[1] Hai Lin,[1] Jinglin Tang,[1] Qihuang Gong,[1,2,3] and Guowei Lu[1,2,3,*]

[1] State Key Laboratory for Mesoscopic Physics, Frontiers Science Center for Nano-optoelectronics & Collaborative Innovation Center of Quantum Matter, School of Physics, Peking University, Beijing 100871, China.
[2] Collaborative Innovation Center of Extreme Optics, Shanxi University, Taiyuan, Shanxi 030006, China.
[3] Peking University Yangtze Delta Institute of Optoelectronics, Nantong 226010, Jiangsu, China.

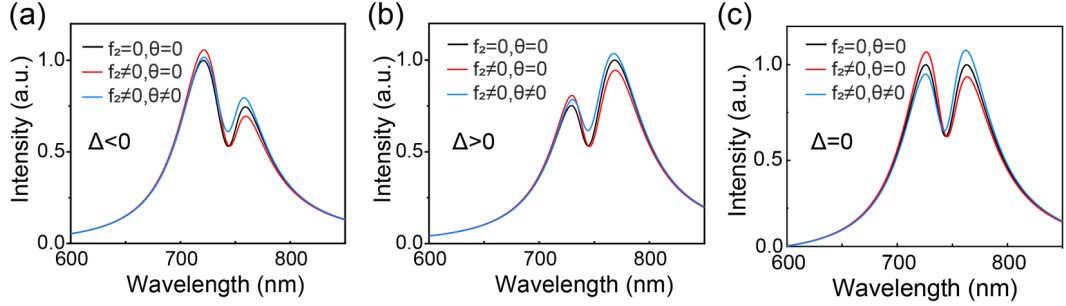

**Figure S1.** Normalized scattering spectra calculated by the model when $E_e = f_2 e^{i\theta} = 0$ (black line), $E_e = 0.06$ (red line) and $E_e = 0.06 e^{i0.9\pi}$ (blue line). The coupling coefficient is 0.03 eV. (a) The resonance wavelength/loss rate of LSPR mode (exciton mode) is at 733 nm (744 nm)/160 meV (48 meV). (b) The resonance wavelength/loss rate of LSPR mode (exciton mode) is at 755 nm (744 nm)/160 meV (48 meV). (c) The resonance wavelength/loss rate of LSPR mode (exciton mode) is at 744 nm (744 nm)/160 meV (48 meV).

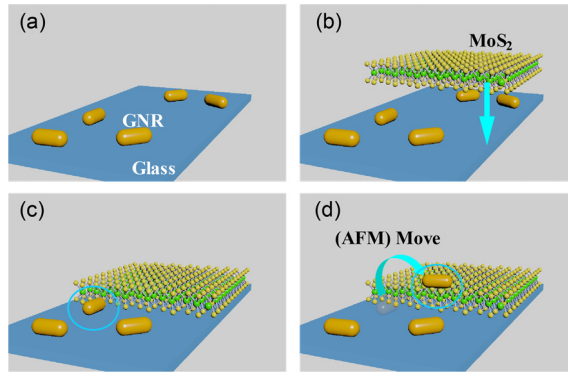

**Figure S2.** (a) The GNRs were synthesized through a seed-mediated wet chemical method [35]. A dilute aqueous solution of GNRs with a suitable concentration was spin-casted onto a glass coverslip to obtain an average spacing of several micrometers. (b) The WSe$_2$ monolayer was transferred onto the glass coverslip. (c) Chose a GNR (blue circle) near the boundary of the WSe$_2$ monolayer (the distance between the GNR and the WSe$_2$ boundary is usually about 1-3 μm). (d) Moving the GNR chosen in (c) from the glass substrate to the WSe$_2$ monolayer using the AFM nanomanipulation.

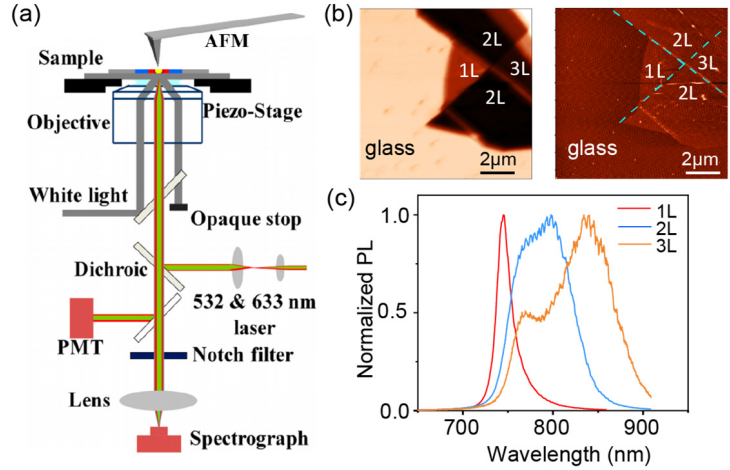

**Figure S3.** (a) Experiment setup. The optical setup is based on the NTEGRA platform (NT-MDT, NTEGRA Spectra, Russia) integrating the white-light dark-field scattering, PL, and AFM techniques [24, 29, 30]. In the experiment, the specific individual GNRs can be determined by AFM images. (b) Optical confocal scanning image (left column) and AFM image (right column) of the GNRs-WSe$_2$ on the glass substrate, with a scale bar of 2 μm. (c) The photoluminescence spectra of monolayer WSe$_2$ (red), bilayer WSe$_2$ (blue), and trilayer WSe$_2$ (orange), measured from the regions labeled as "1L", "2L", and "3L" in (a)-(b). The linewidth of the PL(1L) is about ~42 meV. Excitation is a 532 nm continuous laser.

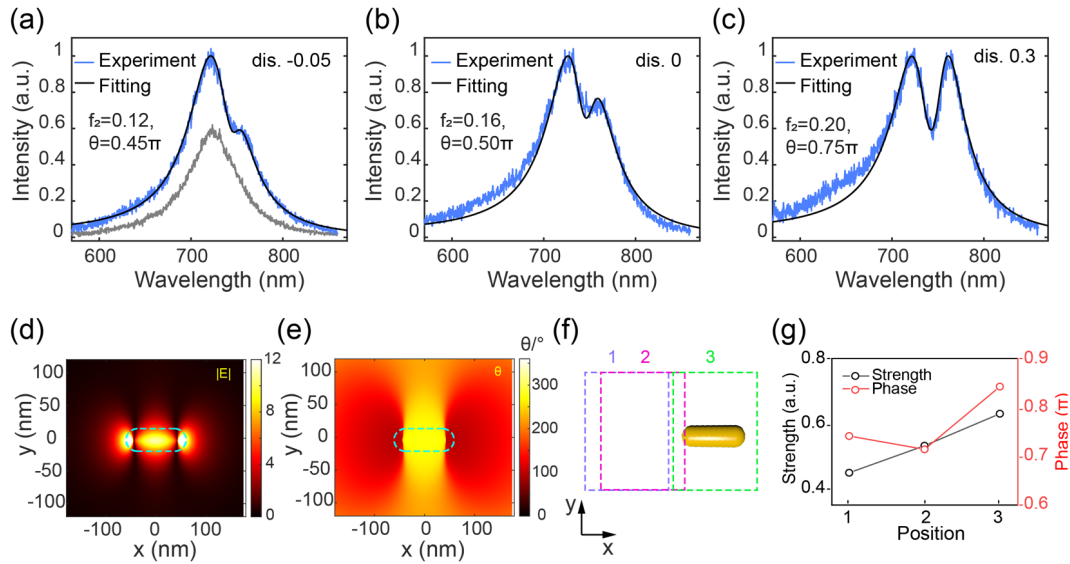

**Figure S4.** Experiment and fitting scattering spectra of the hybrid vary with the distances. (a) Fitting results $f_2 = 0.12, \theta = 0.45\pi$. (b) Fitting results $f_2 = 0.16, \theta = 0.50\pi$. (c) Fitting results $f_2 = 0.20, \theta = 0.75\pi$. The "dis." indicates the separation between the GNR and WSe$_2$.

(d), (e), (f) The strength and phase distribution of the GNR at the bottom surface. The cyan dashed line highlights the geometry of the GNR. (g) The strength and phase vary with different integral areas shown in (f).

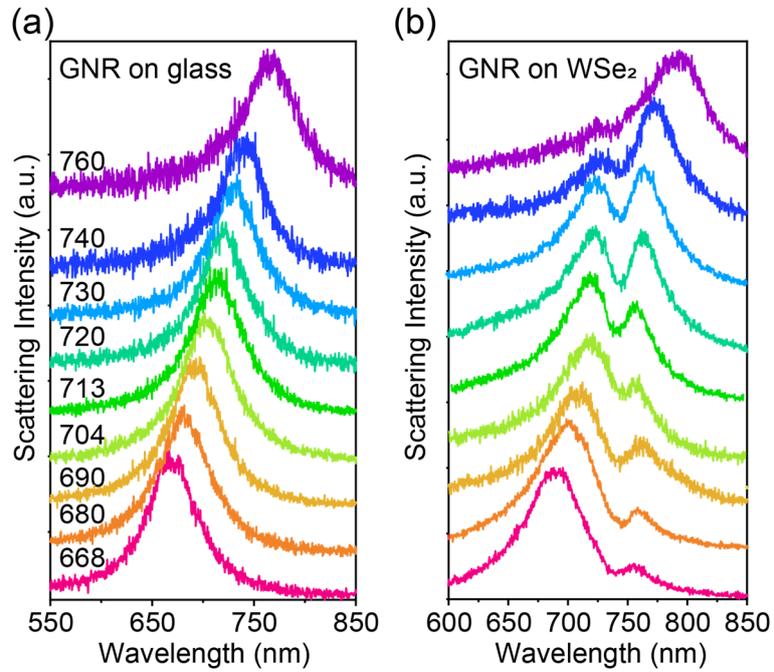

**Figure S5** Coupling of different detuning between the plasmon of GNRs and exciton of monolayer $WSe_2$. (a) Scattering spectra of GNR on the glass substrate. The linewidth of the measured scattering spectra is $130 \pm 20$ meV. (b) Scattering spectra of the hybrids, corresponding to the GNR in (a) was coupling to the monolayer $WSe_2$. The data are corresponding to Fig. 2(e-f). The data in Fig. 2(e-f) in the main text are smoothed.

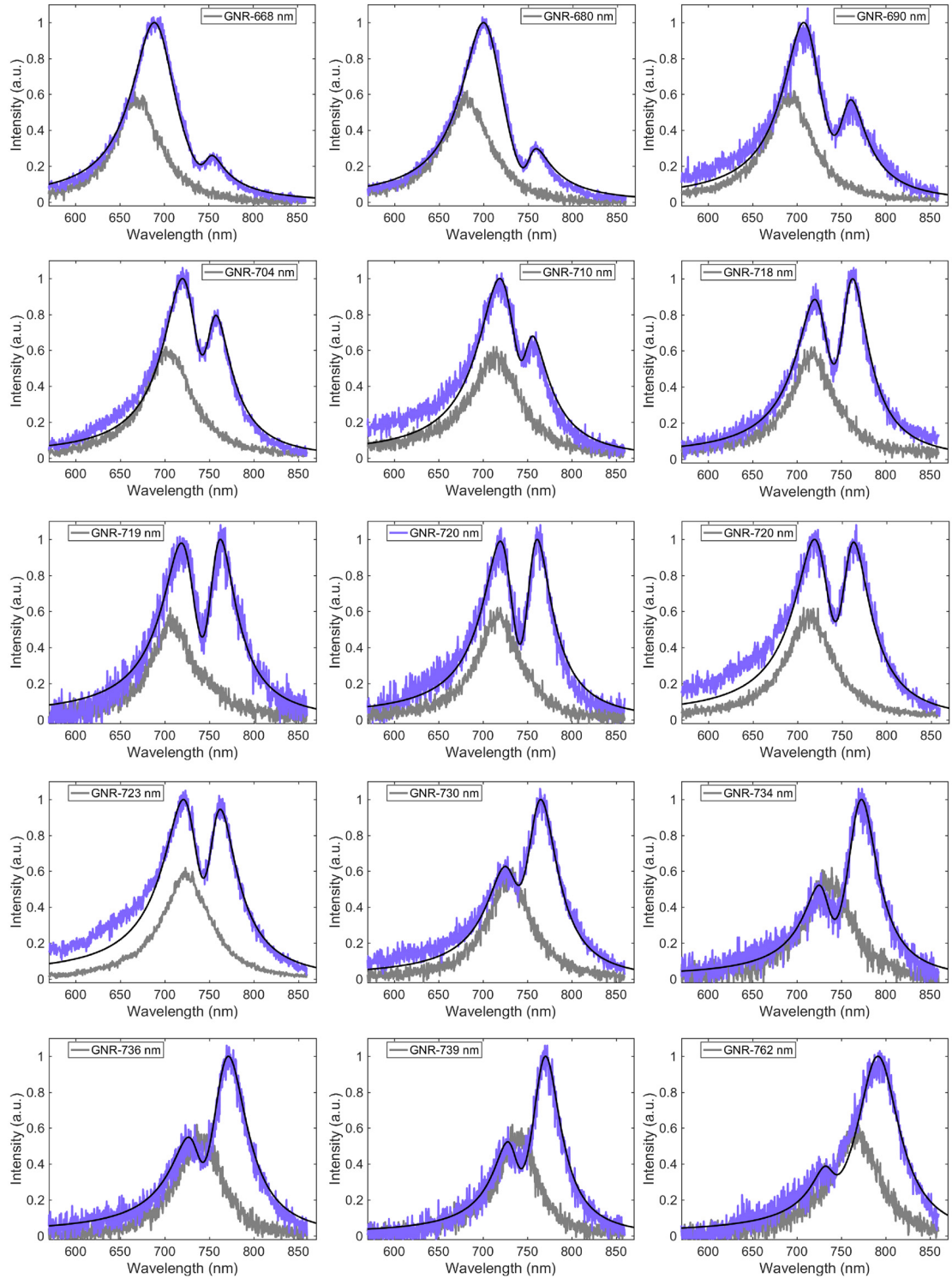

**Figure S6** Experimental scattering spectra (purple line) of GNR-WSe$_2$ hybrids with different detuning and the corresponding fitted results (black line) by the semi-classical model when $f_2 \neq 0$ & $\theta \neq 0$. Experimental scattering spectra of GNR on the glass substrate (gray line). The scattering spectral linewidth of these GNRs is about 130 ± 20 meV.

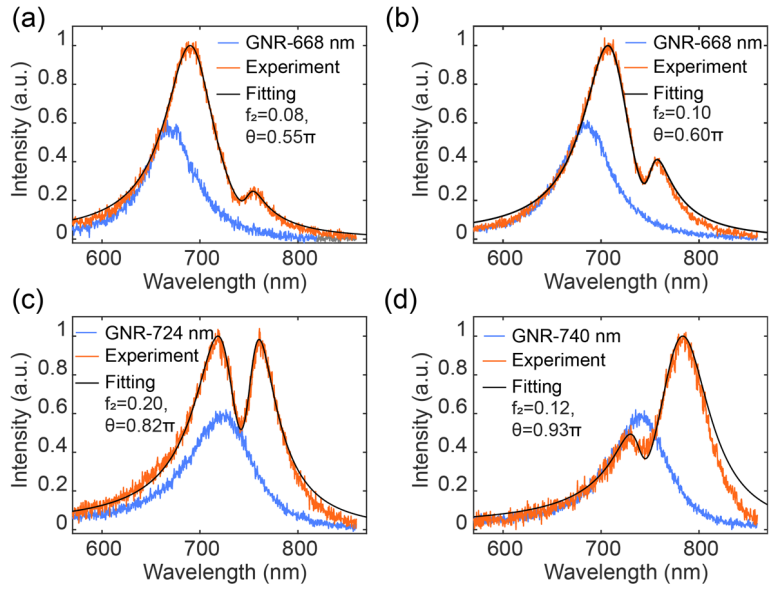

**Figure S7** Experimental and fitted scattering spectra of the hybrid. The scattering spectral linewidth of these GNRs is about 180 ± 20 meV.

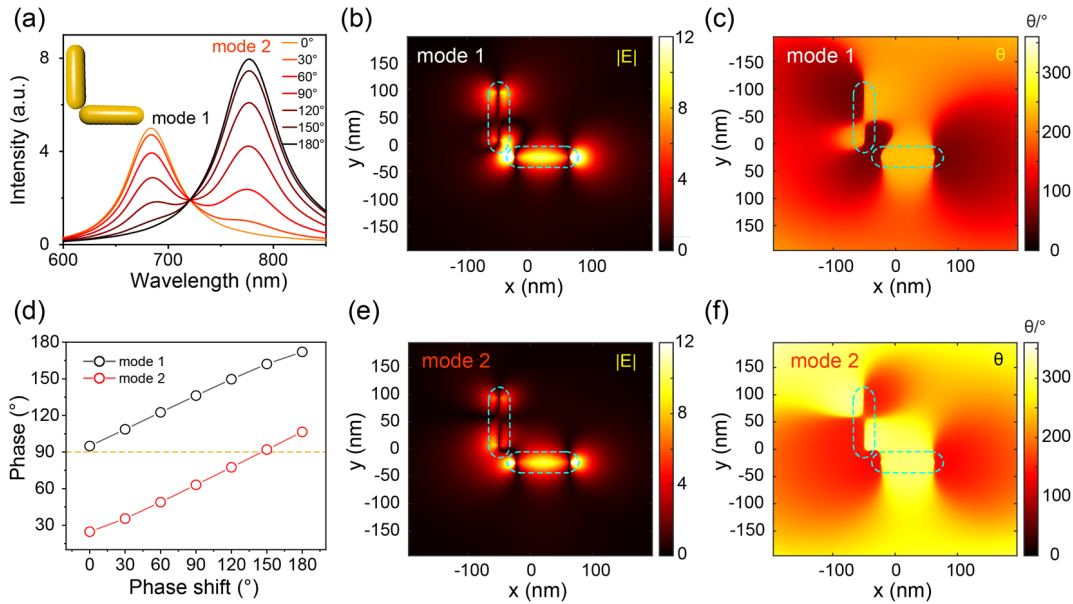

**Figure S8** (a) The spectra of the L-aggregation plasmonic system consisting of two GNR. The excitation light consists of two electromagnetic waves of equal amplitude but differing in phase. The electric near-field (b, e) and local phase (c, f) at the bottom surface of the L-aggregation hybrid at two resonant wavelengths when the excitation phase difference is 90°. The cyan dashed lines highlight the geometry of the GNRs. (d) The modulated local phase under different excitation phase shift. The phase shift indicates the difference between the two electromagnetic waves.

**Modulating the local phase**

The coupling modes theory model (CMTM) usually is used to describe the LSPR behaviors in the plasmonic system. The model also can describe the local phase of a hot spot. For a single GNR with only one LSPR mode, we obtain the scattering spectrum and electric near-field at the bottom surface of GNR excited by the TFSF source. The E-field in Fig. S4(d) is at the resonant wavelength. According to the model,

$$E_{\text{nf}}(\omega, t) \propto \frac{E_{\text{exc}}}{\omega_0 - \omega - i\frac{\gamma_0}{2}} e^{-i\omega t}$$

when $\omega \approx \omega_0$,

$$X(\omega_0, t) \approx e^{i\frac{\pi}{2}} E_{exc} e^{-i\omega t} \tag{S2}$$

It means the local phase at the hot spot is about $\frac{\pi}{2}$, which is consistent in the simulation.

Here, we present a way to modulate the local phase of the hot spot in the plasmonic system. By plasmonic modes coupling, we make a system with two hybrid modes and the hot spots of two hybrid modes are superposition in space. Then, the local phase will be decided by two hybrid modes. In many geometric configurations, the L-aggregation and V-shape can make it. We take the L-aggregation as an example to demonstrate our way.

Two same GNRs are perpendicular to each other. One is along the x-axis and one is along the y axis. The system is symmetric to make sure there is no chirality. TFSF source along the x-axis with a fixed phase and another TFSF source along the y-axis with a variational phase are used to excited. Fig. S8(a) shows the spectra under different excitation phases. Obviously, two hybrid modes can be observed and their proportions change with different excitation phases. We can use CMTM to predict it.

$$E_{\text{nf}}(\omega, t) \propto \left(\frac{E_{exc-1} - E_{exc-2}}{2\left(\omega_0 - \omega - i\frac{\gamma_0}{2} + g\right)} + \frac{E_{exc-1} + E_{exc-2}}{2\left(\omega_0 - \omega - i\frac{\gamma_0}{2} - g\right)}\right) e^{-i\omega t}$$

where $E_{exc-1}$ is with fixed phase, $E_{exc-2}$ is with a variational phase.

When $\omega \approx \omega_0 \pm g$, the local phase can be modulated by setting different excitation phases.

We simulate the *E*-field distribution (Fig. S8(b) and S8(e)) and the local phase (Fig. S8(c)

and S8(f))) at two resonant wavelengths when the excitation phase difference is 90° (circularly polarized). We can observe obvious changes in the local phase at the hot spot, which is consistent in the model. Fig. S8(d) shows the modulated local phase under different excitation phases.

In conclusion, we modulated the local phase at the hot spot of an LSPR mode by introducing another LSPR mode. The spatial superposition leads to interference. And the interference can efficiently modulate the local phase.

**Table S1.** Fitted parameters of the spectra, as shown in the main text Fig. 3(a). The parameters $f_2$ is the driven strength, $\theta$ is the phase, $\omega_p(\omega_e)$ is the resonance frequency of plasmon (exciton) mode, $\gamma_p(\gamma_e)$ is the loss rate of the plasmon (exciton) mode, and g is the coupling coefficient.

| $f_2$ | $\theta/(\pi)$ | $\omega_p$/(nm) | $\gamma_p$/(meV) | $\omega_e$/(nm) | $\gamma_p$/(meV) | g |
|---|---|---|---|---|---|---|
| 0 | 0 | 725 | 166 | 742 | 54 | 0.040 |
| 0.15 | 0 | 723 | 158 | 746 | 54 | 0.040 |
| 0.21 | 0.62 | 723 | 158 | 746 | 54 | 0.047 |

**Table S2.** Fitted parameters of the spectra as shown in the main text Fig. 3(b).

| $f_2$ | $\theta/(\pi)$ | $\omega_p$/(nm) | $\gamma_p$/(meV) | $\omega_e$/(nm) | $\gamma_p$/(meV) | g |
|---|---|---|---|---|---|---|
| 0 | 0 | 748 | 132 | 741 | 50 | 0.039 |
| 0.24 | 0 | 737 | 126 | 746 | 52 | 0.037 |
| 0.245 | 0.87 | 737 | 126 | 746 | 52 | 0.0395 |

**Table S3.** Fitted parameters of the spectra, as shown in the main text Fig. 3(c).

| $f_2$ | $\theta/(\pi)$ | $\omega_p$/(nm) | $\gamma_p$/(meV) | $\omega_e$/(nm) | $\gamma_p$/(meV) | g |
|---|---|---|---|---|---|---|
| 0 | 0 | 762 | 120 | 742 | 52 | 0.037 |
| 0.26 | 0 | 756 | 120 | 746 | 52 | 0.040 |
| 0.28 | 0.88 | 756 | 120 | 746 | 52 | 0.045 |

**Note 1: The finite-difference time-domain (FDTD) Numerical simulation**

The FDTD method, a powerful technique for metallic nanostructures with arbitrary geometries, is employed to calculate the optical responses. The individual gold nanorod (GNR) is modeled as a cylinder capped with hemispheres at each end that was placed on a 500-nm-thick SiO$_2$ layer. Besides, the nanorods with diameter ~ 35 nm and length ~140 ± 30 nm are employed in the simulations. The Drude–Lorentz dispersion model is used for the optical dielectrics of gold, and the refractive indexes of the dielectric media are set to 1.49 for silica, 1.0 for air. The mesh grid is set as 1 nm for other regions. The total scattered-field source is applied. The electric near field (*E*-field) is exported by several steps.

Step 1, simulate a background E-field without GNR as $E_{bg}$.
Step 2, simulate an experiment E-field with GNR as $E_{exp}$.
Step 3, $E_{GNR} = E_{exp} - E_{bg}$.
Step 4, correct the phase of $E_{GNR}$ by setting the phase of exciting light ($E_{bg}$) is 0.